\documentclass[a4paper]{article}

\usepackage{INTERSPEECH2019}
\renewcommand{\thefootnote}{}
\title{Forward-Backward Decoding for Regularizing End-to-End TTS}

\name{Yibin Zheng$^{1,2}$, Xi Wang$^{3}$, Lei He$^{3}$, Shifeng Pan$^{3}$, Frank K. Soong$^{3}$, Zhengqi Wen$^{1}$, Jianhua Tao$^{1,2}$}
\address{
	$^1$Institute of Automation, Chinese Academy of Science, China\\
	$^2$School of Artificial Intelligence, University of Chinese Academy of Science, China \\
	$^3$Microsoft China, Beijing, China}

\email{yibin.zheng@nlpr.ia.ac.cn, xwang@microsoft.com}
\begin{document}

\maketitle
\begin{abstract}
  Neural end-to-end TTS can generate very high-quality synthesized speech, and even close to human recording within similar domain text. However, it performs unsatisfactory when scaling it to challenging test sets. One concern is that the encoder-decoder with attention-based network adopts autoregressive generative sequence model with the limitation of “exposure bias”. To address this issue, we propose two novel methods, which learn to predict future by improving agreement between forward and backward decoding sequence. The first one is achieved by introducing divergence regularization terms into model training objective to reduce the mismatch between two directional models, namely L2R and R2L (which generates targets from left-to-right and right-to-left, respectively). While the second one operates on decoder-level and exploits the future information during decoding. In addition, we employ a joint training strategy to allow forward and backward decoding to improve each other in an interactive process. Experimental results show our proposed methods especially the second one (bidirectional decoder regularization), leads a significantly improvement on both robustness and overall naturalness,  as outperforming baseline (the revised version of Tacotron2) with a MOS gap of 0.14 in a challenging test, and achieving close to human quality (4.42 vs. 4.49 in MOS) on general test.  \let\thefootnote\relax\footnotetext{Samples of synthesized speech are available at https://vancycici.github.io/fbdecode/.} 
\end{abstract}

\section{Introduction}

Recently, with the rapid development of neural network, end-to-end generative text to speech (TTS) models, such as Tacotron and its varieties \cite{Wang2017Tacotron, Shen2017Natural, Chung2018Semi, Wang2017Uncovering} are proposed to simplify traditional TTS pipeline \cite{Taylor2009Text, Black1997Automatically, Zen2009Review, ze2013statistical} with a single neural network. The whole text sequence and corresponding frame-level acoustic features could be effectively learned in a unified network, and with the help of WaveNet \cite{Oord2016WaveNet} like neural vocoder, the quality and naturalness of synthesized audios are greatly improved, and even comparable with human recordings within similar domain text. The end-to-end TTS model is based on encoder-decoder framework \cite{Bahdanau2014Neural}, which has been widely adopted for sequence generation tasks, such as end-to-end speech recognition \cite{Chen2010Improving} and neural machine translation (NMT) \cite{Hassan2018Achieving, xia2017deliberation}. This encoder-decoder framework also brings appealing properties, such as little requirement for feature engineering or prior domain knowledge, unified network instead of fragment components, flexible transformation and etc, making it easier to construct a high quality and expressive TTS system \cite{Skerry2018Towards, Wang2018Style, Stanton2018Predicting, Jia2018Transfer}.

However, in the framework of end-to-end TTS, the decoder is an autoregressive structure which will prevent the usage of global or future output information during training and inference. That is, while generating current frame, one can only leverage the generated frames but not the future frames un-generated so far. This forward-decoding process will constrain the usage of global or future information. What's worse, a tiny mistake made early could quickly be amplified and propagated alongside the sequence (so-called exposure bias problem \cite{bengio2015scheduled}). Further, such issue may become severe when the test sequences are not matched with training data, and would become much severer as the sequence length increases. Various attention mechanisms \cite{li2018close, vaswani2017attention, chorowski2015attention} have been incorporated to find a more accurate mapping from encoder to decoder. However, due to the constrain of autoregressive generation, the decoding bias problem could not be well addressed only by attention mechanism itself. In  \cite{bengio2015scheduled}, the author provides a simple way to address the ``exposure bias", namely scheduled sampling \cite{bengio2015scheduled}, which is done by combing the predicted probability with ground-truth label. To better leverage the global or future information, there are also some attempts in speech recognition \cite{Chen2010Improving} and NMT \cite{Hassan2018Achieving, Liu2016AgreementOT}, which try to improve the performance by integrating the predicted probability from forward and backward decoding sequences. However, regarding TTS outputs are continuous space, these methods  \cite{Hassan2018Achieving, bengio2015scheduled, Liu2016AgreementOT} could not be directly employed in TTS.

Actually, the whole utterance is known during training, and with the help of estimation of future information, the generating sequence would be much more robust. Then the accumulated error would be fixed soon and would not be propagated for a long way. The assumption is based on the agreement between forward and backward decoding while generating sequence, and thus the attention should be more accurate. In this paper, we propose two novel methods to realize this forward-backward decoding sequence. 1) Train two directional models, namely L2R (which generates targets from left-to-right) and R2L (which generates targets from right-to-left), and iteratively update the model by a novel training objective with additional consideration of minimizing both directional model divergence at the same time, in which the latter one servers as a measure of exposure bias of the currently evaluated model. 2) With shared encoder, train bi-directional decoders with their own attentions. By adding regularization term for forward-backward decoders to the training objective, the forward hidden states are forced to be well close to the backward ones. Thus, it could encourage the hidden representations of a unidirectional decoder to embed some useful information about the future.  For both two methods, a joint training strategy is proposed to make forward and backward decoders improve each other in an interactive update process. Furthermore, the backward network could be omitted during inference, leading to the model that ideally does not introduces any latency and does not add any computation compared to the standard unidirectional end-to-end TTS model.

\section{Proposed Methods}

To better leverage the global or future information as well as to alleviate the exposure bias problem, we describe in depth the two proposed methods that integrate forward and backward decoding sequences here.

\subsection{Model regularization by bidirectional agreement}
\label{2-1}

To predict future as well as to deal with the exposure bias problem, we try to maximize the agreement between the generated spectrograms from L2R and R2L end-to-end TTS models, and divide the training objective into two parts: one for the standard \(L_2\) loss of each model, and the other for regularization terms that indicate the divergence of L2R and R2L models based on the current model parameters. 
\subsubsection{Model regularization}
\label{2-1-1}
Given a  text sequence \(x=(x_1,x_2,...,x_T)\) and its target mel spectrograms \(y=(y_1,y_2,...,{y_{T'}})\), let \(P(y|x;\overrightarrow{\theta})\) and \(P(y|x;\overleftarrow{\theta})\) be L2R and R2L model, in which \(\overrightarrow{\theta}\) and \(\overleftarrow{\theta}\) are corresponding model parameters. Specifically, L2R model can be decomposed as \(P(y|x;\overrightarrow{\theta})=\prod_{t=1}^{T'}P(y_t|y_{<t},x;\overrightarrow{\theta})\), which means L2R  model adopts previous targets \(y_1,...,{y_{t-1}}\) as history to predict current target \(y_t\) at each step \(t\), while R2L model similarly can be decomposed as \(P(y|x;\overleftarrow{\theta})=\prod_{t=T'}^{1}P(y_t|y_{>t},x;\overleftarrow{\theta})\) and employs future targets \(y_{t+1},...,{y_{T'}}\) as history to predict current target  \(y_t\) at each step \(t\). 

Since L2R and R2L models are different chain decompositions of the same output sequence, output sequence of these two models should be ideally identical:
\begin{equation}
\setlength{\abovedisplayskip}{2pt}
\setlength{\belowdisplayskip}{2pt}
P(y|x;\overrightarrow{\theta})=P(y|x;\overleftarrow{\theta})
\label{eq1}
\end{equation}

However, the above equation will not hold if these two models are optimized separately by standard loss of each model. To satisfy this constrain, we introduce a \(L_2\) regularization term into the training objective. For L2R model, the new training objective now becomes:
\begin{equation}
\setlength{\abovedisplayskip}{2pt}
\setlength{\belowdisplayskip}{2pt}
L(\overrightarrow{\theta})=\sum_{n=1}^{N}L_2(y_{\overrightarrow{\theta}}^{(n)},y^{(n)})+\lambda\sum_{n=1}^{N}L_2(y_{\overrightarrow{\theta}}^{(n)},y_{\overleftarrow{\theta}}^{(n)})
\label{eq2}
\end{equation}
where \(\lambda\) is a hyper-parameter for regularization term. The regularization term will guide the training process to reduce the disagreement between L2R and R2L model.

Considering the symmetry of L2R and R2L model, L2R model can also serve as the discriminator to punish bad generation candidates generated from R2L. Based on the above, L2R and R2L model can act as helper systems for each other in a joint training process.

Though both L2R and R2L are needed at training time, we could use either L2R or R2L model during inference. This leads to equal amount of computations, compared to standard unidirectional model (at inference time).


\begin{figure}[h]
	
	\vspace{-12pt}
	\centering
	\includegraphics[scale=0.4]{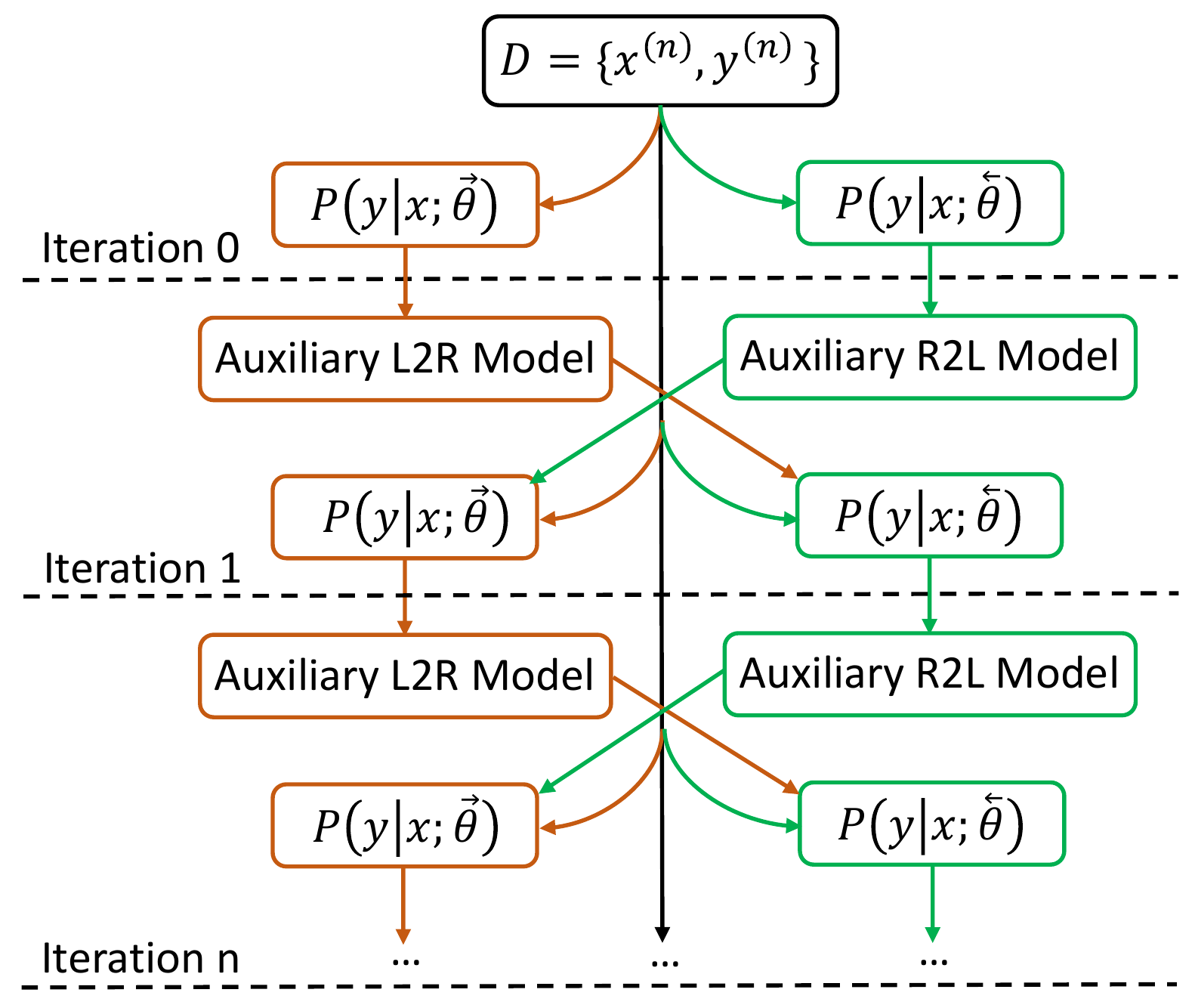}
	\vspace{-5pt}
	\caption{Illustration of joint training of L2R \& R2L model.}
	\label{fig:fig}
	\vspace{-18pt}
\end{figure}

\subsubsection{Joint training for model regularization of L2R \& R2L}
\label{sssec:subsubhead}

To simultaneously optimize these two models, we design a novel training algorithm and the overall training objective is defined as the sum of objectives in both directions:
\begin{equation}
\setlength{\abovedisplayskip}{2pt}
\setlength{\belowdisplayskip}{2pt}
L({\theta})=L(\overrightarrow{\theta})+L(\overleftarrow{\theta})
\label{eq3}
\end{equation}

As illustrated in Fig.~\ref{fig:fig}, the whole training process contains two major steps: pre-training and joint-training. Firstly, we pre-train both L2R and R2L models with standard loss of each end-to-end TTS model. Next, based on the pre-trained models, we jointly optimize L2R and R2L models with an iterative process. In each iteration, we fix R2L model and use it as an auxiliary helper system to optimize L2R models with Eq.\ref{eq2}, and at the same time, we fix L2R model and use it as an auxiliary helper system to optimize R2L model. This training process could be carried out to obtain further improvements because after each iteration both L2R and R2L model are expected to be improved with regularization method.

\subsection{Bi-directional decoder regularization}
\label{2-2}
The method in Sec.~\ref{2-1} is operated in the model level, which is similar to data augmentation (since we have to generate pseudo \(<text, audio> \) pairs from reversed model). However, for the data generation is very time consuming and not effective.  We consider the second method to further implement forward-backward decoding. Since the unidirectional decoders tend to be more accurate at their early decoding steps, which leads to a strategy that forward and backward decoder can be integrated to boost the final performance. As shown in  Fig.~\ref{fig:fig2}, in which the forward decoder is trained with a subtask of backward decoding by sharing a single encoder and vice versa, aiming at a regularization effect of optimization for both direction decoder.

\subsubsection{Bi-directional decoder regularization}
\label{2-2-1}
\begin{figure}[h]
	\vspace{-10pt}
	\centering
	\includegraphics[width=\linewidth]{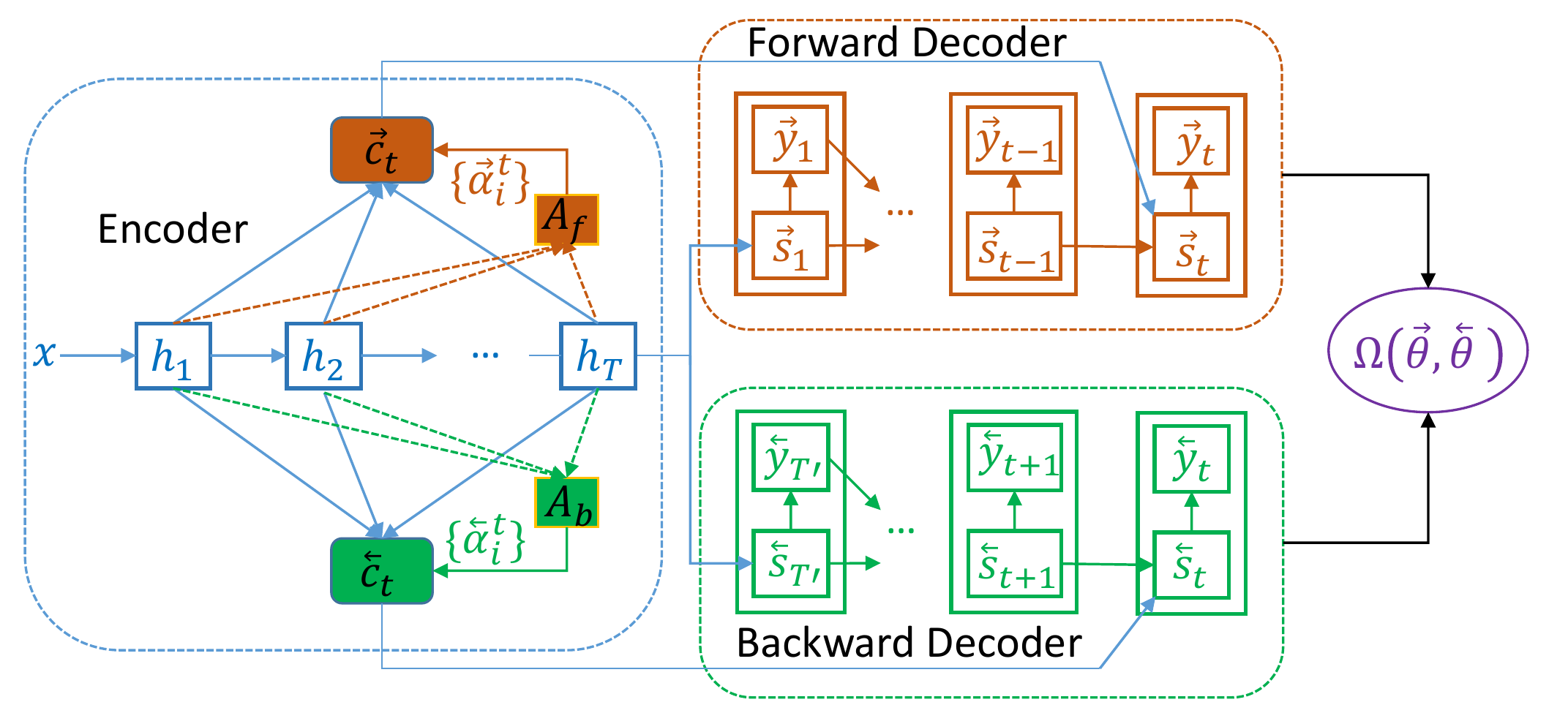}
	\vspace{-15pt}
	\caption{Bi-direction decoder regularization. Blue, orange, and green parts indicate encoder, forward-decoder and backward-decoder, respectively.}
	\label{fig:fig2}
	\vspace{-20pt}
\end{figure}

In  Fig.~\ref{fig:fig2}, the shared encoder represents an input text sequence \(x=(x_1,x_2,...,x_T)\) into hidden representation of \textit{\textbf{h}}, and the forward decoder computes, at each time step \(t\), a new hidden state \(\overrightarrow{s}_t\) in the following way:
\begin{equation}
\setlength{\abovedisplayskip}{2pt}
\setlength{\belowdisplayskip}{2pt}
h_t=encoder(h_{t-1}, x_t)
\label{eq4}
\end{equation}
\begin{equation}
\setlength{\abovedisplayskip}{2pt}
\setlength{\belowdisplayskip}{2pt}
\overrightarrow{s}_t=decoder_{\overrightarrow{\theta}}(\overrightarrow{s}_{t-1}, y_{t-1}, \overrightarrow{c}_t)
\label{eq5}
\end{equation}
where \(c_t\) is  the  context  vector  calculated  by  a  location-sensitive attention mechanism \cite{Shen2017Natural}.

The forward states summarize the information about current and past elements of the sequence. Similarly, it is very convenient to also process the sequence in the reversed time order, and compute backward states \(\overleftarrow{s}_t\) similarly to Eq.\ref{eq5}:
\begin{equation}
\setlength{\abovedisplayskip}{2pt}
\setlength{\belowdisplayskip}{2pt}
\overleftarrow{s}_t=decoder_{\overleftarrow{\theta}}(\overleftarrow{s}_{t+1}, y_{t+1}, \overleftarrow{c}_t)
\label{eq7}
\end{equation}

The backward states summarize the information about current and future elements of the sequence. To encourage the hidden states of a unidirectional decoder to embed some useful information about the future, we add a regularization term, as highlighted in  Fig.~\ref{fig:fig2}. The idea is to penalize forward hidden states \(\overrightarrow{s}_t\) that are distant from the backward ones \(\overleftarrow{s}_t\). With this regard, one can add a regularization term that encourage the network to minimize the \(L_2\) distance between forward and backward hidden states:
\begin{equation}
\setlength{\abovedisplayskip}{2pt}
\setlength{\belowdisplayskip}{2pt}
\Omega=\frac{1}{T'}\sum_{t=1}^{T'}\left\|\overrightarrow{s}_t-\overleftarrow{s}_t\right\|^2
\label{eq8}
\end{equation}

The total objective to be minimized thus becomes a weighted sum of the standard loss plus the regularization term:
\begin{equation}
\setlength{\abovedisplayskip}{2pt}
\setlength{\belowdisplayskip}{2pt}
\tilde{L}(\theta)=L(\overrightarrow{\theta})+L(\overleftarrow{\theta})+\lambda\Omega(\overrightarrow{\theta}, \overleftarrow{\theta})
\label{eq9}
\end{equation}
where \(\lambda\) is a hyper-parameter controlling the importance of the penalty term, and \(\tilde{L}(\theta)\) is the total loss.

Note the backward states are needed only at training time. During inference, the part of the model computing backward states can be omitted. This leads to the architecture particularly suitable for practical use, since it requires exactly the same amount of computations needed for standard unidirectional decoder (at inference time). Another remarkable aspect of this technique is that Eq.\ref{eq7} is based on the backward states \(\overleftarrow{s}_t\), that provide a summary of the full future part of the mel spectrograms sequence. This means that our method could capture not only short-term future dependencies, but also long-term ones.

\subsubsection{Joint-training of bi-directional decoder regularization}
In practice, it is hard to train the whole network from scratch jointly since the revered decoder cannot provide useful enough information at the beginning of training. To simultaneously optimize bi-directional decoders, we design a novel training algorithm, which is similar to that in  Fig.~\ref{fig:fig}. The whole training contains two major steps: pre-training and joint-training. First, we pre-train both forward and backward jointly without adding the regularization term. Next, based on the pre-trained bi-directional decoders, we jointly optimize forward and backward decoders with an iterative process using Eq.\ref{eq8}. In each iteration, we fix backward decoder and use it as a helper to optimize forward decoder, and vice versa.

\section{Experiments}

\label{sec:majhead}
In this section, we conduct experiments to evaluate our proposed methods a 20-hour, 16kHz, 16bit speech corpus, which is recorded by a professional enUS female speaker. All the subjective tests are evaluated by at least 10 native judges from Microsoft crowdsourcing UHRS (Universal Human Relevance System) platform.

\subsection{Model details}
\label{ssec:subhead}
For our baseline, we use similar architecture and hyper parameters as Tacotron2 \cite{Shen2017Natural} except a few details. We test our proposed methods with both phoneme and character inputs for end-to-end model training. The decoder outputs 80-channel mel spectrograms, one frame at a time. We train a WaveNet vocoder conditioned on mel spectrograms with a constant learning rate of \(10^{-4}\), and use it for waveform generation. Our WaveNet is a smaller one than \cite{Shen2017Natural}, with only 12 dilated layers \cite {arik2017deep}. 

The L2R and R2L model have the same architecture with the baseline model. For joint training algorithm of model regularization,  we find that this method is very time consuming (since we have to generate pseudo \(<text, audio> \) pairs from reversed model in each iteration) and thus only 1 full iteration is performed to be evaluated here. Besides, we find that the R2L model gets comparable results with the L2R model, thus only the result of the L2R model is
reported.

As for bi-directional decoder regularization, the architectures of encoder and decoder are kept the same with the baseline model. We perform 5 full iterations for the joint training of bidirectional decoder. Besides, we find that the forward decoder gets comparable results with the backward decoder, thus only the result of the forward decoder is reported.

Hyper-parameter \(\lambda\) in  Eq.\ref{eq9} is set as 1.0 and all the models are trained with 4 Nvidia Tesla P100 using Adam optimizer \cite{kingma2014adam}, with initial rate of \(10^{-3}\) and decays exponentially after 50000 steps. 

\subsection{Evaluation}
\label{ssec:subhead}
\subsubsection{Evaluation on out-of-domain text}
\begin{figure}[h]
	\vspace{-10pt}
	\centering
	\includegraphics[width=\linewidth]{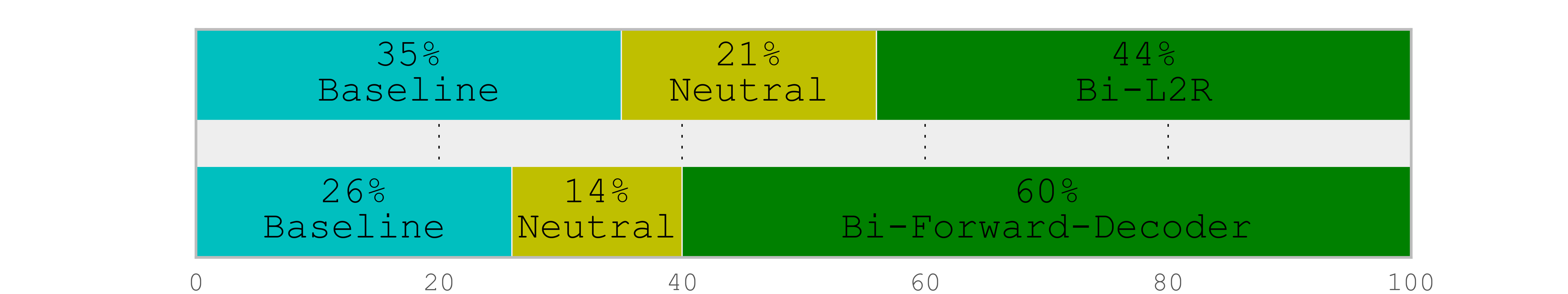}
	\vspace{-14pt}
	\caption{The results of  AB preference test by using character as input, with confidence level of 95\% and \(p\)-value \(<0.0001\).}
	\label{fig:fig3}
	\vspace{-15pt}
\end{figure}

Firstly, we evaluate the proposed methods on out-of-domain text. In this section, we collect a large challenging test set, including news, long facts, foreign words, difficult abbreviation, numbers, letters, strange text normalization (TN) extension, etc. For subjective evaluation, we randomly select 50 test utterances from the challenging test set.  We first conduct two AB preference tests on these 50 audios pairs generated by baseline and our proposed two methods respectively.  In this AB preference test, we use character as the input for the end-to-end TTS training. The AB preference test results are shown in Fig.~\ref{fig:fig3}, in which \textbf{Baseline} denotes the end-to-end baseline model (revised version of Tacotron2 \cite{Shen2017Natural}), \textbf{Bi-L2R} denotes the joint-trained L2R model introduced in Sec.~\ref{2-1}, and \textbf{Bi-Forward-Decoder} represents the joint-trained forward decoder by adding hidden state regularization that introduced in Sec.~\ref{2-2}. We find that both two proposed methods consistently receive more preferences than the baseline model. These results confirm that the proposed two regularization methods by introducing agreement between forward and backward decoding sequence can help handling exposure bias problem and improving the synthesized audios' naturalness. Specifically, instead of combining backward model during inference, both two proposed methods utilize the intrinsic connection between forward and backward decoding models to guide the learning process. The forward and backward decoding models are expected to adjust in disagreement cases and then the exposure bias problem of them can be alleviated. And among these models, the \textbf{Bi-Forward-Decoder} gains the most preferences (60\%). It shows from judges' comments that \textbf{Bi-Forward-Decoder} tends to generate more natural and human-like speech, meanwhile it reduces the probability of running into pronunciation difficulties, e.g., when handling names or out of vocabulary (OOV) words.  Therefore, model \textbf{Bi-Forward-Decoder} is selected for further more experiments. 

We further train models \textbf{Bi-Forward-Decoder} and \textbf{Baseline} with phoneme as input. Together with  character-based models, 5-point mean opinion score (MOS) tests are conducted.  Tab.~\ref{tab:example3} shows the out-of-domain evaluation results of different models. It's noticed that the proposed model \textbf{Bi-Forward-Decoder} performs better than baseline model \textbf{Baseline} no matter what kind of input representation is used.  And a much more obvious advantages of our proposed method could be observed when using character as input. This could be explained by that,  compared  with phoneme-based models, character-based models often suffer from  pronunciation issues caused by grapheme-to-phoneme,  and our proposed method is helpful to correct such issues since it could use the global information of an utterance to give more proper pronunciation.  And among these systems, the \textbf{Bi-Forward-Decoder} (with phoneme as input) obtains the best performance with a MOS of 4.26.

\begin{table}[h]
	\vspace{-5pt}
	\caption{The MOS of different models in out-of-domain test, with confidence level of 95\%.}
		\vspace{-5pt}
	\label{tab:example3}
	\centering
	\begin{tabular}{ l c c c}
		\toprule
		Inputs & Character  & Phoneme  \\
		\hline
		Baseline    & 3.77  & 4.12             \\
		\textbf{Bi-Forward-Decoder}         & \textbf{4.07}   & \textbf{4.26}             \\
		Recording         & 4.48   & 4.48            \\
		\bottomrule
	\end{tabular}
	\vspace{-8pt}
\end{table}


Secondly, we select another 100 test utterances from the large challenge test set for intelligibility test. This set is much more challenging with many abbreviation, letters, numbers, repeated words, OOV, long sentences, etc, which may lead the attention collapses. The number of words in these 100 test utterances is 2261. Each test utterance will be evaluated by two native judges, and both unintelligible and unnatural word will be marked. In this task, we count the following two matrices and the results are shown in Tab.~\ref{tab:example} and Tab.~\ref{tab:example1}. These two matrices include: case level intelligible rate which means the proportion of the cases without any words marked as “unintelligible” in a test case; case level natural rate which means the proportion of the cases without any word marked as “unintelligible” or “unnatural” (such as emphasis on the wrong syllables or words, or unnatural pitch) in a test case. It's observed that both the proposed character-based and phoneme-based \textbf{Bi-Forward-Decoder} significantly improve the model's intelligibility and naturalness, with an absolute improvement of  4.0\% and 4.5\%, respectively for character-based model, 1.6\% and 1.8\% , respectively for phoneme-based model (which is not that significant as character-based model, but still beneficial).
\setlength{\tabcolsep}{1.2mm}
\begin{table}[t]
	\caption{Case level intelligible rate and natural rate of character-based models.}
		\vspace{-5pt}
	\label{tab:example}
	\centering
	\begin{tabular}{lcc}
		\toprule
		Models      & Baseline & \textbf{Bi-Forward-Decoder}                \\
		\midrule
		case level intelligible rate                       & 95.5\% & \textbf{99.5\%}             \\
		case level natural rate                       & 93.5\% & \textbf{{98.0\%}}             \\
		\bottomrule
	\end{tabular}
	\vspace{-5pt}
\end{table}

\begin{figure}[h]
	\vspace{-8pt}
	\centering
	\includegraphics[width=\linewidth]{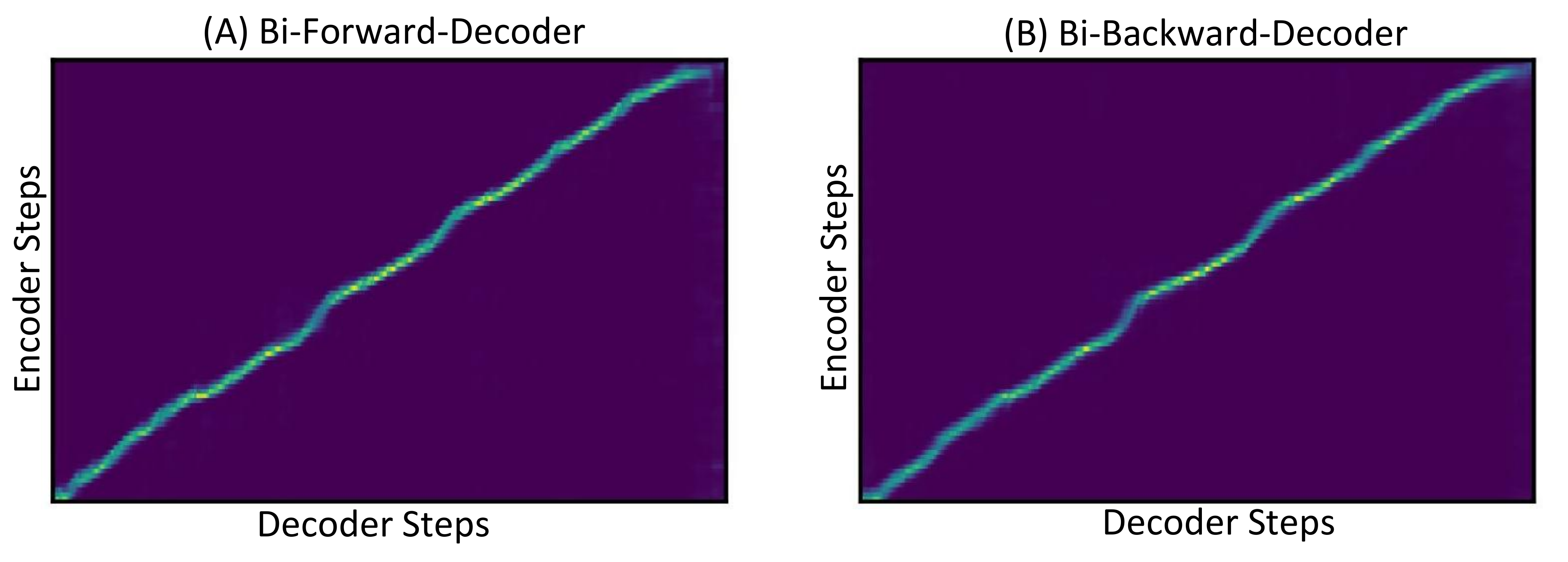}
	\vspace{-14pt}
	\caption{Attention alignments on a test utterance.}
	\label{fig:fig4}
	\vspace{-20pt}
\end{figure}

\begin{table}[t]
	\caption{Case level intelligible rate and natural rate of phoneme-based models.}
		\vspace{-5pt}
	\label{tab:example1}
	\centering
	\begin{tabular}{lcc}
		\toprule
		Models      & Baseline & \textbf{Bi-Forward-Decoder}                \\
		\midrule
		case level intelligible rate                       & 96.8\% & \textbf{98.4\%}             \\
		case level natural rate                       & 95.6\% & \textbf{{97.8\%}}             \\
		\bottomrule
	\end{tabular}
	\vspace{-10pt}
\end{table}
Apart from the quantitative analysis, we also give a case study to better understand how the regularization method works. We show the alignments of the joint-trained \textbf{Bi-Forward-Decoder} and \textbf{Bi-Backward-Decoder} with the same text, as shown in Fig.~\ref{fig:fig4}. As we can see, the alignment of forward and backward decoder is almost the same after regularization, which means the forward and backward decoder reach a high degree of agreement.

\subsubsection{Evaluation on in-domain text}
Apart from the out-of-domain evaluation, we also evaluate the proposed method on relative in-domain text, like appropriate text length and content. In this section, only the model \textbf{Bi-Forward-Decoder} (which performs best on out-of-domain text) is included for comparison. We randomly select 50 test utterances (not included in training set)  from our internal dataset. Tab.~\ref{tab:examplein} shows the in-domain evaluation results of different models. We find that the  both the proposed character-based and phoneme-based \textbf{Bi-Forward-Decoder} perform better than the baseline model \textbf{Baseline}, with a gap of 0.14 and 0.06 in MOS, respectively. This trend is similar as that on out-of-domain evaluation, which further confirms the effectiveness of the proposed regularization method.  Meanwhile, it's noticed that phoneme-based model \textbf{Bi-Forward-Decoder} obtains the best performance with a MOS of 4.42, which is quite close to recording (4.49). Examination of judges' comments also show that \textbf{Bi-Forward-Decoder} performs much better on overall prosody, sounds more expressive, stable and clearer than \textbf{Baseline}.

\begin{table}[h]
	\vspace{-2pt}
	\caption{The MOS of different models in in-domain test, with confidence level of 95\%.}
	\vspace{-6pt}
	\label{tab:examplein}
	\centering
	\begin{tabular}{ l c c c}
		\toprule
		Inputs & Character  & Phoneme  \\
		\hline
		Baseline      & 4.10  & 4.36             \\
		\textbf{Bi-Forward-Decoder}         & \textbf{4.24}   & \textbf{4.42}             \\
		Recording         & 4.49   & 4.49            \\
		\bottomrule
	\end{tabular}
	\vspace{-13pt}
\end{table}


\section{Conclusions}
\label{sec:print}

In this paper, we propose two efficient regularization training approaches to the end-to-end TTS framework, aiming to improve the robustness of the model. Relying on the optimization of the agreement between forward and backward decoding sequence, the forward decoder could be better optimized with both global and future information of the output, thus gains much more stable, expressive results but without any additional computation during inference. Experimental results demonstrate that our methods (especially the bidirectional decoder regularization method), achieves a significant improvement on robustness and naturalness on both in-domain and out-of-domain evaluation. However, due to the model characteristic of encoder-decoder based framework, it still suffers from incorrect issues on a few extremely rare cases. We will keep on this work to improve the stableness of end-to-end TTS .

\vspace{-3pt}
\section{Acknowledgements}
The author would like to thank Shujie Liu and Fei Tian from Microsoft research with fruitful discussion.

\vfill
\pagebreak

\bibliographystyle{IEEEtran}

\bibliography{mybib}


\end{document}